\documentclass[%
 reprint,
superscriptaddress,
showkeys,
 amsmath,amssymb,
 aps,
pre, 
onecolumn
]{revtex4-2}

\usepackage{graphicx}
\usepackage[caption=false]{subfig} 

\usepackage{dcolumn}
\usepackage{bm}

\usepackage[
top=3cm, 
left=3cm,
right=3cm,
bottom=4cm,
]{geometry}

\newcommand{\ee}{\end{equation}} 
\newcommand{\be}{\begin{equation}}

\usepackage[mathscr]{euscript}  
\usepackage{xcolor}  

\usepackage{tcolorbox}
\usepackage{comment}
\usepackage{float}

\usepackage{hyperref}
\hypersetup{
    colorlinks=true,
    linkcolor=black,
    citecolor=black,
    filecolor=black,
    urlcolor=black,
    }

\makeatletter
\newsavebox{\@brx}
\newcommand{\llangle}[1][]{\savebox{\@brx}{\(\m@th{#1\langle}\)}%
  \mathopen{\copy\@brx\kern-0.5\wd\@brx\usebox{\@brx}}}
\newcommand{\rrangle}[1][]{\savebox{\@brx}{\(\m@th{#1\rangle}\)}%
  \mathclose{\copy\@brx\kern-0.5\wd\@brx\usebox{\@brx}}}
\makeatother

\begin{document}

\preprint{ApS/123-QED}

\title{Thermodynamic work of partial resetting}

\author{Kristian St\o{}levik Olsen}
\thanks{kristian.olsen@hhu.de}
\affiliation{Institut für Theoretische Physik II - Weiche Materie, Heinrich-Heine-Universität Düsseldorf, D-40225 Düsseldorf, Germany}

\author{Deepak Gupta}
\thanks{phydeepak.gupta@gmail.com}
\affiliation{Department of Physics, Indian Institute of Technology Indore, Khandwa Road, Simrol, Indore-453552, India}
\affiliation{Nordita,  Royal Institute of Technology and Stockholm University,  Hannes Alfvéns väg 12, 23, SE-106 91 Stockholm, Sweden}

\begin{abstract}
Partial resetting, whereby a state variable $x(t)$ is reset at random times to a value $a x (t)$, $0\leq a \leq 1$, generalizes conventional resetting by introducing the resetting strength $a$ as a parameter.  Partial resetting generates a broad family of non-equilibrium steady states (NESS) that interpolates between the conventional NESS at strong resetting ($a=0$) and a Gaussian distribution at weak resetting ($a \to 1$). Here, such resetting processes are studied from a thermodynamic perspective,  and the mean cost associated with maintaining such NESS are derived. The resetting phase of the dynamics is implemented by a resetting potential $\Phi(x)$ that mediates the resets in finite time. By working in an ensemble of trajectories with a fixed number of resets, we study both the steady-state properties of the propagator and its moments. The thermodynamic work needed to sustain the resulting NESS is then investigated. We find that different resetting traps can give rise to rates of work with widely different dependencies on the resetting strength $a$. Surprisingly, in the case of resets mediated by a harmonic trap with otherwise free diffusive motion, the asymptotic rate of work is insensitive to the value of $a$. For general anharmonic traps, the asymptotic rate of work can be either increasing or decreasing as a function of the strength $a$, depending on the degree of anharmonicity. Counter to intuition, the rate of work can therefore in some cases increase as the resetting becomes weaker $(a\to 1)$ although the work vanishes at $a=1$. Work in the presence of a background potential is also considered. Numerical simulations confirm our findings. 
\end{abstract}

\pacs{Valid pACS appear here} 
\maketitle


\section{Introduction}

Over the past decade, stochastic resetting has become a vibrant area of research within non-equilibrium statistical physics. Resetting randomly interrupts a { \color{black}system's evolution} and re-starts it again according to some pre-defined scheme \cite{evans2011diffusion, evans2011optimal}. These recurrent perturbations to the system { \color{black}drive} it away from equilibrium and into a non-equilibrium steady state (NESS). For this reason, resetting has become { \color{black} an arena} for exploring non-equilibrium phenomena that are accessible both in theory and in experiments. 

Various forms of resetting can be found in both natural and man-made situations. A wide range of extreme events bear the hallmarks of resetting, namely a large and sudden fluctuation whereby the system (approximately) starts afresh. In the study of animal behavior, resetting is a common strategy of foraging whereby search phases are interrupted by recurrent returns to a central location \cite{bell1990central,wiltschko2023animal}.  In computer science it is well-known that restarts can aid in shortening the completion time of various algorithms \cite{luby1993optimal,montanari2002optimizing,blumer2022stochastic}. 

In statistical physics, resetting became popular after the work of Evans and Majumdar a little over a decade ago  \cite{evans2011diffusion, evans2011optimal}. Since then a vast range of resetting schemes have been invented. Within the conventional paradigm, a system state  (typically { \color{black} the position of a} particle) is instantaneously reset to its initial condition at constant rate. This paradigm has been applied to many different situations, including diffusion in the presence of potentials or other forms of heterogeneities \cite{Pal_PRE, ray2020diffusion,KSO2023,sandev2022heterogeneous}, active matter \cite{evans2018run,kumar2020active,sar2023resetting,santra2020run,bressloff2020occupation}, and search processes \cite{evans2011diffusion, bressloff2020search, reuveni2016optimal,pal2017first,chechkin2018random, pal2019first, ahmad2019first,ahmad2022first,tucci2022first,singh2022first,de2020optimization,besga2020optimal}. Variations in the resetting scheme itself has also been considered, for example by considering more complex inter-resetting times \cite{pal2016diffusion,shkilev2017continuous,eule2016non, nagar2016diffusion,radice2022diffusion}. For a review, see Ref.~\cite{evans2020stochastic}.

Another version of resetting that is gaining more attention in recent years is partial resetting (also sometimes called proportional resetting). Here the position is not reset completely to some reference position, but is rather reset by an amount proportional to its current value. For a scalar parameter $a \in [0,1]$ that determines the strength of the reset, the position is in this scheme scaled to $a x(t)$ at each resetting event. Such partial resets share a similarity to so-called Markovian growth-collapse models which have been studied in the past \cite{boxma2006markovian,lopker2011hitting,privault2022moments}, and have connections to stress-releases taking place during earthquakes \cite{vere1988variance,zheng1991application}. Partial resets are also believed to be relevant to catastrophic events in population dynamics \cite{hanson1981logistic,gripenberg1983stationary}. In the statistical physics literature, partial resetting has appeared only recently. Ref. \cite{dahlenburg2021stochastic} considers resets by a random amplitude, which roughly corresponds to partial resets with random strengths $a$. Ref. \cite{pierce2022advection} studied partial resets of velocity in an advection-diffusion model, and calculates both moments and steady-state properties. Further developments were made in Ref. \cite{tal2022diffusion}, where the case of diffusion and drift-diffusion was considered in some detail. More recently, full time-dependent analysis for Markov processes was carried out in Ref. \cite{di2023time}.

Over recent years, the thermodynamics of conventional resetting has become fairly well understood \cite{fuchs2016stochastic,pal2017integral,pal2021thermodynamic,busiello2020entropy, Deepak2022_work,olsen2023thermodynamic,mori2023entropy}. Any resetting scheme that gives rise to a NESS is expected to have associated with it a non-zero rate of entropy production or thermodynamic work. The thermodynamics of instantaneous resetting is somewhat artificial since no physical implementation of resetting is instantaneous. A natural framework that attributes clear thermodynamic interpretations to the various elements of resetting is intermittent fluctuating potentials. Here a resetting trap is turned on, presumably at a random instance of time, in order to bring a particle to a prescribed position. This resetting scheme has been studied both in theory \cite{gupta2020stochastic,mercado2020intermittent,santra2021brownian,mercado2022reducing, Gupta_2021_SR,alston2022non,xu2022stochastic} and in recent experimental implementations of resetting using optical tweezers \cite{goerlich2023experimental,tal2020experimental,besga2020optimal}. { \color{black} Other types of costs that do not necessarily correspond to thermodynamic quantities has also been considered recently \cite{sunil2023cost,sunil2024minimizing}.}

Here we study the stochastic thermodynamics of partial resets, with resetting mediated by a confining trap. The resetting strength $a$ is known to strongly affect the steady state characteristics \cite{tal2022diffusion}, and hence it is relevant to understand the associated thermodynamic cost at different resetting strengths{ \color{black}, which until now remains unexplored}. By considering an ensemble of trajectories where exactly an integer number of resets have been completed, we calculate steady state properties valid for a wide range of systems and use these insights to calculate the mean rate of thermodynamic work, and study its dependence on the resetting strength parameter $a$. { \color{black} The results may be of relevance to experimental resetting protocols, or to design swift state-to-state transitions between non-equilibrium steady states as discussed in Ref. \cite{goerlich2024resetting}. Additionally, the resetting strength $a$ allows a continuous interpolation between conventional resetting and no resetting, with interesting thermodynamic properties along the way.}

This paper is organized as follows. Section \ref{sec:model} presents the main ingredients in the model, the equations of motion, and the details of the resetting scheme. Section \ref{sec:moments} studies various properties of the system, such as steady-state propagators and moments in various scenarios. Section \ref{sec:work} proceeds to compute the thermodynamic work needed to implement a protocol with exactly $n$ resets. The case of a harmonic and anharmonic quartic resetting trap is treated in detail, and expected behavior{ \color{black}s} for other resetting trap shapes are discussed. We also consider the work in the presence of a potential in the exploration phase between resets. Finally, section \ref{sec:concl} offers a concluding discussion and potential outlooks. 

\begin{figure}
    \centering
    \includegraphics[width = \textwidth]{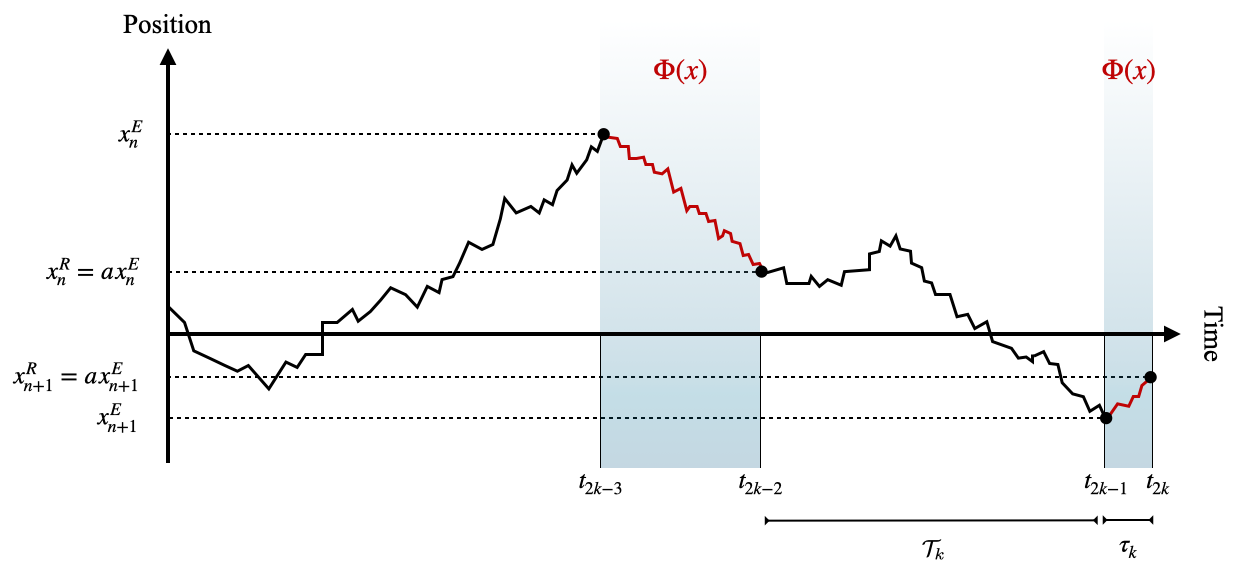}
    \caption{Partial resetting mediated by a intermittent potential $\Phi(x)$. Exploration trajectories (black) in the $n$'th epoch end at points $x^E_n$, with positions after reset obeying $x^R_{n} = a x^E_n$. Resetting trajectories (red) last for a duration equal to the first passage time from $x^E_n$ to $x^R_{n}$.}
    \label{fig:scheme}
\end{figure}

\section{Model and setup}\label{sec:model}
We consider an overdamped Brownian particle { \color{black} in one dimension} that in the absence of resetting obeys the equation of motion

\begin{equation}\label{eq:underlying}
    \frac{d x(t)}{dt} = \sqrt{2D} \xi(t) + \mu F[x(t)]\ ,
\end{equation}
where $\xi(t)$ a Gaussian white noise of zero mean $\langle \xi(t)\rangle=0$ and unit variance $\langle \xi(t)\xi(t')\rangle= \delta(t-t')$. We will assume that the force $ F[x(t)]$ originates from a potential landscape $V(x)$. In the above equation of motion, the diffusivity $D$ is related to the mobility $\mu$ through the standard Einstein relation $D = k_BT\mu$, with $k_B$ Boltzmann's constant which we will set to unity, and $T$ the temperature of the heat bath.

To implement resetting, we consider a resetting trap $\Phi(x)$, which is intermittently switched on and off. Assuming that $\Phi(x)$ is confining $\Phi''(x) > 0$, the particle will be drawn towards the trap minimum as long as the trap is active. We refer to the periods without a resetting trap as \emph{exploring phases}, and the periods where the resetting trap is active as \emph{resetting phases}. The particle starts in an exploration phase. We denote the times at which the dynamics changes from exploring to resetting or \emph{vice versa} as $t_i$, and note that $\mathcal{T}_k \equiv t_{2k-1} - t_{2k-2}$ is the duration of the $k$'th exploration phase, and $\tau_k \equiv t_{2k}-t_{2k-1} $ the duration of the $k$'th resetting phase [see Fig. (\ref{fig:scheme})], for $k\geq 1$.

Under the effect of such a fluctuating potential, we extend Eq. (\ref{eq:underlying}) to 
\begin{equation}
    \frac{d x(t)}{dt} = \sqrt{2D}\xi(t) - \lambda(t) \mu  \frac{\partial \Phi(x)}{\partial x}  - [1-\lambda(t)] \mu  \frac{\partial V(x)}{\partial x}\ , 
\end{equation}
where we have introduced an external control parameter $\lambda(t)$ that is zero in the exploration phase and takes the value $1$ in the resetting phase:
{ \color{black}\begin{equation}\label{eq:lambda}
    \lambda(t) = 
\left\{
	\begin{array}{ll}
		0  & \mbox{if } t \in (t_{2k-2}\ ,~t_{2k-1})\ , \\
		1  & \mbox{if } t \in (t_{2k-1}\ ,~t_{2k})\ .
	\end{array}
\right.
\end{equation}}
In this way, the total time-dependent potential landscape the particle experiences can be written
\begin{eqnarray}\label{eq:totpot}
    V_{tot} (x,\lambda) =  \lambda(t)   \Phi(x)  + [1-\lambda(t)]   V(x)\ .
\end{eqnarray}
{ \color{black} This is motivated by recent experiments on resetting where the potential can be rapidly switched from a sharp to a shallow state \cite{goerlich2023experimental}.
We here note that if instead we considered a protocol where the background potential $V(x)$ was always present and only $\Phi(x)$ was switched on and off, the same framework would hold with the replacement $\Phi(x)\to \Phi(x)+V(x)$.}

Associated with the above external control protocol{ \color{black}, Eq. (\ref{eq:totpot}),} is the thermodynamic work 
\cite{seifert2012stochastic}
\begin{eqnarray}\label{eq:thework}
    w[\{x_t\}] = \int_0^t dt' \frac{\partial V_{tot}[x(t'),\lambda(t')]}{\partial \lambda(t')} \frac{d \lambda(t')}{d t'}\ ,
\end{eqnarray}
which depends functionally on both the particles trajectory and the protocol. { \color{black}
The first law of thermodynamics relates the work $w$ to the total change in energy $\Delta E$ and the dissipated heat $q$ through $E[\{x_t\}] = w[\{x_t\}] - q[\{x_t\}]$, where the dissipated heat is given by \cite{seifert2012stochastic}
\begin{eqnarray}\label{heat-eqn}
    q[\{x_t\}] = \int_0^t dt' F_{tot}[x(t'),\lambda(t')]\circ \dot x(t')\ .
\end{eqnarray}
Here $F_{tot}[x,\lambda] = -\partial_x V_{tot}(x,\lambda)$ is the force acting on the particle, and the integral~\eqref{heat-eqn} is interpreted according to the Stratonovich convention. For overdamped systems, the overall change in energy $\Delta E$ is simply the change in potential energy, which is bounded in time. Hence, we expect that at late times, the rate of growth of the (average) work and (average) dissipated heat are equal. The entropy production in the medium associated with the dissipation $S_{m}[\{x_t\}] = q[\{x_t\}]/T$ was studied in Ref. \cite{mori2023entropy} for diffusion with conventional resetting ($a=0$). For further comparisons between thermodynamical quantities associated with conventional resetting, see Ref. \cite{olsen2023thermodynamic}.
}

The main quantity of interest in this { \color{black} paper is the thermodynamic work, Eq. (\ref{eq:thework})}, which is studied in Section \ref{sec:work}. We will for the sake of generality consider a resetting protocol where the durations of the exploration phases $\mathcal{T}_k$ are distributed with a general density $\psi(\mathcal{T})$. { \color{black} We will throughout assume that this density has well-defined moments.} The resetting phases however, have a duration that is much more complex, as it relates to the first passage time from a current (random) location $x$ to $a x$. For $a=0$, a similar problem was studied in Ref. \cite{Deepak2022_work}. As we will see, one can often avoid solving this first-passage problem by working in an ensemble where trajectories exactly end at the terminal points of the exploration or resetting phases.

\section{Results in the fixed-$n$ ensemble}\label{sec:moments}
While the partial resetting protocol gives rise to a rather complex time-evolution, we can get around some of these troubles by utilizing the fact that the dynamics has a natural temporal discretization. Indeed, the state at times $t_k$ at which the dynamics switches from exploring to resetting (or \emph{vice versa}) gives us much information regarding the behavior of the system, although not fully time-resolved. This is similar to the approach taken to study sharp resetting at periodic time intervals in Ref. \cite{tal2022diffusion}. We refer to this approach as the fixed-$n$ ensemble, similar to ensembles of paths considered in similar systems in the past \cite{olsen2023thermodynamic,mori2023entropy,mori2021condensation,mori2020universal,hartmann2020convex,singh2022mean}. The main motivation for considering such discrete dynamics in this paper, however, is thermodynamics. The work associated with  maintaining the resetting process involves an energetic cost at each instant of potential switching \cite{Deepak2022_work,olsen2023thermodynamic}. Hence, knowledge regarding the systems state at these instants will allow us to calculate the mean work in section \ref{sec:work}.

\subsection{Propagator at the end of the exploration phases}

{ \color{black} 
In the exploration phase [under the influence of $V(x)$], $G_E(x,t|y)$ is the propagator of the particle. Then, the} 
probability density for the particle position restricted to the end of the $n$'th exploration phase 
reads
\begin{equation}\label{eq:rhoE1}
    \rho_n(x^E_n) = \int d x^R_{n-1} \int d\mathcal{T}\psi(\mathcal{T})  G_E(x^E_n,\mathcal{T}|x^R_{n-1}) \pi_{n-1}(x^R_{n-1})\ ,
\end{equation}
where $\pi_{n-1}(x^R_{n-1})$ is the distribution of positions at the end of the $(n-1)$'th resetting phase. Here the particle propagates with the exploration propagator $G_E(x,\mathcal{T}|x^R_{n-1}) $ for a time $\mathcal{T}$, distributed according to $\psi(\mathcal{T})$, initialized from the position after the previous resetting $x^R_{n-1}$. The distribution of $x^R_{n}$ in turn, can be obtained directly from $x^R_{n} = a x^E_n$. Using basic transformation properties of probability densities we simply have
\begin{eqnarray}
       \pi_{n}(x^R_{n}) &= \frac{1}{a}\rho_n(x^R_{n}/a)\ . \label{eq:rhoRfinal}
\end{eqnarray}
{ \color{black} We here note that intermittent potentials that act on the particle for random durations (see Ref.~\cite{alston2022non}) also gives rise to a type of partial resetting, however not with constant strength $a$. For general fluctuating potentials, Eq.~(\ref{eq:rhoRfinal}) becomes more involved, and the partial resetting constraint $x^R_{n} = a x^E_n$ here offers a significant simplification~\cite{olsen2023thermodynamic}. Using Eq.~(\ref{eq:rhoRfinal}) together with Eq.~(\ref{eq:rhoE1}) we can}  write a closed integro-recursive relation for the density $\rho_n(x)$, which reads
\begin{equation}\label{eq:rhoEfinal}
    \rho_n(x^E_n) =  \int d[x^E_{n-1}] \int d\mathcal{T}\psi(\mathcal{T})  G_E(x^E_n,\mathcal{T}|a x^E_{n-1})  \rho_{n-1}\left(x^E_{n-1}\right)\ .
\end{equation}
Eq.~(\ref{eq:rhoEfinal}) together with Eq.~(\ref{eq:rhoRfinal}) then equips us with complete information about the particles positions after the exploration and resetting phases as a function of ``discrete time'' $n$. Notice that all dependence on the resetting trap $\Phi(x)$ has disappeared - hence all expectation values in the fixed-$n$ ensemble will not depend on the details of the resetting { \color{black} trap} except for the reset strength $a$. In section \ref{sec:work} however, we will see that the mean thermodynamic work has properties that strongly depends on the chosen resetting trap { \color{black}$\Phi(x)$}.

\subsection{Free exploration}\label{sec:freeexpl}
When there is no exploration potential $V(x) =0$, the above analysis simplifies and we can study the steady-state (i.e. large-$n$) properties of $\rho_n(x^E_n)$. This has properties similar to the fully time-resolved steady state studied for example in Ref. \cite{tal2022diffusion}. At large $n$ we can drop the subscript in Eq. (\ref{eq:rhoEfinal}), and we have
\begin{equation}
    \rho(x^E_n) =  \int d[x^E_{n-1}] \int d\mathcal{T}\psi(\mathcal{T})  G_E(x^E_n,\mathcal{T}|a x^E_{n-1})  \rho \left(x^E_{n-1}\right)\ .
\end{equation}
Since the process is spatially homogeneous{ \color{black},} we expect that the propagator in the exploration phase satisfies
\begin{eqnarray}
    G_E(x^E_n,\mathcal{T}|a x^E_{n-1}) = G_E(x^E_n-a x^E_{n-1},\mathcal{T})\ .
\end{eqnarray}
This brings Eq. (\ref{eq:rhoEfinal})  into a convolution form. By performing a Fourier transform, we can write
\begin{eqnarray}
    \hat \rho(k) &=  \int d\mathcal{T}\psi(\mathcal{T})  \hat G_E(k,\mathcal{T})  \hat \rho\left(a k\right)\ , \\
    & = \prod_{j=0}^\infty  \left[ \int d\mathcal{T}\psi(\mathcal{T})\hat G_E(a^jk,\mathcal{T}) \right]\ , \label{eq:gen}
\end{eqnarray}
where we iterated the equation indefinitely to find the solution. Following \cite{tal2022diffusion} we note that $ \hat{\rho}(k) $ has the natural interpretation as the Fourier transform of the density of the steady state position  $X = \sum_{j=0}^\infty z_j$, where { \color{black} the $z_j$'s are independent random variables} related to the traveled distance in the $j$'th exploration phase, and { \color{black} follows a distribution whose Fourier transform reads} 
\begin{equation}
    \hat p_j(k) \equiv  \int d\mathcal{T}\psi(\mathcal{T})\hat G_E(a^jk,\mathcal{T})\ . 
\end{equation}
Assuming a zero-mean process, the variance of $p_j(x)$ can be found simply by applying a double derivative with respect to $k$ and letting $k\to0$. This gives
\begin{equation}\label{eq:var}
    \Sigma_j^2 =  \int d\mathcal{T}\psi(\mathcal{T})a^{2j} \sigma^2(\mathcal{T})\ ,
\end{equation}
where $\sigma^2(\mathcal{T})$ is the variance of the underlying process calculated from the exploration propagator $G_E(x,t|y)$, { \color{black}and it is independent of the resetting strength $a$}. { \color{black} Since for free exploration the variables $z_j$ are independent, } the variance of the variable $X$ is { \color{black} simply the sum of the variances from Eq. (\ref{eq:var})}
\begin{equation}\label{eq:ss_sig}
    \Sigma^2(a) \equiv \sum_{j=0}^\infty \Sigma_j^2 = { \color{black} \int d\mathcal{T}\psi(\mathcal{T}) \sigma^2(\mathcal{T}) \left(\sum_{j=0}^\infty a^{2j}\right) = } \frac{\int d\mathcal{T}\psi(\mathcal{T}) \sigma^2(\mathcal{T})}{1-a^{2}}  \ . 
\end{equation}
{ \color{black} Here we assumed a strict inequality $a<1$ when performing the geometric sum. } This clearly diverges as  { \color{black} the strength, $a$, approaches $1$}. Define the scaled observable $Z = X/\Sigma(a)$, and denote its probability density $P(Z;a)$; this is nothing but our end-of-exploration density $\rho(x)$ re-scaled with the { \color{black}standard deviation, i.e.,}
\begin{equation}\label{eq:Pz}
    {P}(z;a) = \Sigma(a) {\rho}( z \Sigma(a))\ .
\end{equation}
In Ref.~\cite{tal2022diffusion} a transition in the steady state density for Brownian motion with Poissonian resetting was observed as the resetting strength $a$ was varied. In the case of a freely diffusive Brownian particle, strong resetting $a\to 0$ resulted in a Laplace distribution, while for weak resetting $a\to 1$ the steady state was found to approach a Gaussian { \color{black}distribution}. The same transition holds in the present case for arbitrary $\psi(\mathcal{T})$, namely 
\begin{eqnarray}
    \lim_{a\to 1} {P}(z;a) = e^{- k^2/2}
\end{eqnarray}
Hence, while a steady state appears at all $a<1$, this steady state becomes a Gaussian { \color{black}distribution} with infinite width in this limit. We elaborate on this in the appendix.

\subsubsection{Variance and kurtosis.} 
Moments for any $n$ can be obtained directly from Eq. (\ref{eq:rhoEfinal}). For the second moment, we multiply Eq. (\ref{eq:rhoEfinal}) with $(x^E_n)^2$ and integrate, resulting in 
\begin{equation}
    \langle (x^E_n)^2 \rangle=  \int d x^E_{n-1} \int d\mathcal{T}\ \psi(\mathcal{T})\   \langle (x^E_n)^2| a x^E_{n-1} \rangle_E(\mathcal{T})\   \rho_{n-1}\left(x^E_{n-1}\right)\ , \label{sec-mom}
\end{equation}
where $\langle (x^E_n)^2| x^E_{n-1} \rangle_E(\mathcal{T})$ is the moment calculated using the conditional exploration propagator $G_E(\cdot)$. { \color{black} For a pure diffusion process with diffusivity $D$, the} propagator is Gaussian in the case of free exploration, we have the second moment
\begin{eqnarray}
    \langle (x^E_n)^2| a x^E_{n-1} \rangle_E(\mathcal{T}) = 2 D \mathcal{T} + a^2 ( x^E_{n-1})^2\ . \label{cond-sec-mom}
\end{eqnarray}
Inserting { \color{black}Eq.~\eqref{cond-sec-mom} in Eq.~\eqref{sec-mom}} gives 
\begin{equation}
    \langle (x^E_n)^2 \rangle = 2 D \langle \mathcal{T}\rangle + a^2 \langle( x^E_{n-1})^2\rangle\ . \label{sec-mom-2}
\end{equation}
Using as initial condition $x(0) = 0$, the { \color{black}above solution~\eqref{sec-mom-2}} reads
\begin{eqnarray}\label{eq:secmomfree}
    \langle (x^E_n)^2 \rangle = 2 D \langle \mathcal{T}\rangle  \frac{1-a^{2n}}{1-a^2}\ .
\end{eqnarray}

In the limit of weak resetting $a\to 1$, we find $\langle (x^E_n)^2 \rangle = 2 D n \langle \mathcal{T}\rangle $, which is nothing but the second moment of a Brownian particle at time $n \langle \mathcal{T}\rangle$. For $a <1$ the steady state value $2 D \langle \mathcal{T}\rangle/(1-a^2)$ is obtained by letting $n\to \infty$ and using the fact that $a^{2n}\to 0$. The same result can be obtained directly from Eq. (\ref{eq:ss_sig}) { \color{black}, simply by observing that for Brownian motion $\sigma^2(\mathcal{T} ) = 2 D \mathcal{T}$, so that $\int_0^\infty d\mathcal{T} \psi(\mathcal{T}) 2 D \mathcal{T} = 2 D \langle\mathcal{T}\rangle$}.

For the fourth moment, we multiply Eq. (\ref{eq:rhoEfinal}) with $(x^E_n)^4$ and integrate, resulting in 
\begin{equation}
    \langle (x^E_n)^4 \rangle=  \int d[x^E_{n-1}] \int d\mathcal{T}\psi(\mathcal{T})  \langle (x^E_n)^4| a x^E_{n-1} \rangle_E(\mathcal{T})  \rho_{n-1}\left(x^E_{n-1}\right)\ . \label{four-moment}
\end{equation}
Since the propagator in the exploration phase is Gaussian, we have the fourth moment
\begin{eqnarray}
    \langle (x^E_n)^4| a x^E_{n-1} \rangle_E(\mathcal{T}) = 12 D^2 \mathcal{T}^2 + 12 a^2 D \mathcal{T}    ( x^E_{n-1})^2 +  a^4  ( x^E_{n-1})^4\ . \label{four-moment-con}
\end{eqnarray}
Inserting { \color{black}Eq.~\eqref{four-moment-con} in Eq.~\eqref{four-moment}} gives 
\begin{equation}
    \langle (x^E_n)^4 \rangle = 12 D^2 \langle \mathcal{T}^2\rangle  + 12 a^2 D \langle  \mathcal{T}\rangle \langle ( x^E_{n-1})^2\rangle +  a^4 \langle ( x^E_{n-1})^4\rangle\ .  \label{four-mom-2}
\end{equation}
In principle,~{ \color{black}Eq.~\eqref{four-mom-2}} can be solved recursively as was done for the second moment. However, we can also directly calculate { \color{black}its} steady state value. In the steady state we can { \color{black}drop} the subscript $n$. Using previous results for the steady state second moment{ \color{black}~\eqref{eq:secmomfree}}, we have 
\begin{equation}
    \langle (x^E)^4 \rangle = 12 D^2 \langle \mathcal{T}^2\rangle  +  a^2  \frac{24 (D \langle \mathcal{T}\rangle)^2 }{1-a^2} +  a^4 \langle ( x^E)^4\rangle \ . \label{four-mom-3}  
\end{equation}
Solving { \color{black}above equation~\eqref{four-mom-3}} gives the steady state value 
\begin{equation}\label{eq:fourth}
   \langle (x^E)^4 \rangle = \frac{1}{1-a^4}\left[12 D^2 \langle \mathcal{T}^2\rangle  +  a^2  \frac{24 (D \langle \mathcal{T}\rangle)^2 }{1-a^2} \right]\ .
\end{equation}

From the steady state second { \color{black}\eqref{eq:secmomfree}} and fourth moment~{ \color{black}\eqref{eq:fourth}} we can calculate the kurtosis of the steady state. Since the process has zero mean, we simply have
\begin{eqnarray}
    \textnormal{Kurt}(x;a) = \frac{ \langle (x^E)^4 \rangle }{ \langle (x^E)^2 \rangle ^2}\ .
\end{eqnarray}
The aforementioned transition to a Gaussian { \color{black}distribution} in the weak resetting limit, implies that $ \textnormal{Kurt}(x;1)  = 3$, which can be confirmed using the above moment values. For $a\to 0$ on the other hand, we find 
\begin{eqnarray}
    \textnormal{Kurt}(x;0) = 3\frac{\langle \mathcal{T}^2\rangle}{\langle \mathcal{T}\rangle^2}\ ,
\end{eqnarray}
signifying that the non-Gaussianity is determined by the statistics of the durations of the exploration phase. For the exponential case $\psi(\mathcal{T}) = r \exp(-r \mathcal{T})$ we have $\langle \mathcal{T}\rangle = 1/r$ and $\langle \mathcal{T}^2\rangle = 2/r^2$, giving $\textnormal{Kurt}(x;0) = 6$ as in Ref. \cite{tal2022diffusion}.

\subsection{Variance in a harmonic exploration potential}
While the exact steady state is not easy to obtain for a non-zero exploration potential $V(x)$, we can obtain exact values for the moments. We consider a harmonic potential $V(x) = \frac{1}{2}{ \color{black}\kappa} x^2$. As before, we multiply Eq. (\ref{eq:rhoEfinal}) with $(x^E_n)^2$ and integrate, which results in 
\begin{equation}\label{sec-2-mom-1}
    \langle (x^E_n)^2 \rangle=  \int d[x^E_{n-1}] \int d\mathcal{T}\psi(\mathcal{T})  \langle (x^E_n)^2| a x^E_{n-1} \rangle_E(\mathcal{T})  \rho_{n-1}\left(x^E_{n-1}\right)\ .
\end{equation}
 For the second moment calculated in the exploration phase, we have 
\begin{eqnarray}\label{sec-2-mom-2}
    \langle (x^E_n)^2| a x^E_{n-1} \rangle_E(\mathcal{T}) = D \tau_{\rm rel} \left(1-e^{-2 \mathcal{T} /\tau_{\rm rel}}\right) + a^2 e^{-2 \mathcal{T} /\tau_{\rm rel}}  (x^E_{n-1})^2\ ,
\end{eqnarray}
where $\tau_{\rm rel} = \gamma/{ \color{black}\kappa}$ is the relaxation timescale of the potential. We find
\begin{equation}\label{sec-2-mom3}
    \langle (x^E_n)^2 \rangle =  D \tau_{\rm rel} \left(1-\tilde\psi (2/\tau_{\rm rel})\right) + a^2 \tilde\psi (2/\tau_{\rm rel})   \langle (x^E_{n-1})^2 \rangle\ ,
\end{equation}
with tildes denoting Laplace transforms. This equation~{ \color{black}\eqref{sec-2-mom3}} can be solved recursively, subject to the initial condition that a particle that at time $t=0$ starts at the origin. This results in 
\begin{eqnarray}\label{sec-2-mom4}
    \langle (x^E_n)^2 \rangle &=  \frac{D \tau_{\rm rel} (\tilde\psi(2/\tau_{\rm rel})  -1) \left[\left(a^2 \tilde\psi(2/\tau_{\rm rel})  \right)^n-1\right]}{1-a^2 \tilde\psi(2/\tau_{\rm rel})  } \ .
\end{eqnarray}
In the steady state, { \color{black}the right-hand side of Eq.~\eqref{sec-2-mom4}} approaches a constant, which is obtained by letting $n\to \infty$:
\begin{eqnarray}\label{eq:Vx2}
   \lim_{n\to \infty} \langle (x^E_n)^2 \rangle &=  D \tau_{\rm rel} \frac{1- \tilde\psi(2/\tau_{\rm rel})}{1-a^2\tilde\psi(2/\tau_{\rm rel})}\ .
\end{eqnarray}
A couple of things are worth noting:
\begin{itemize}
    \item [i)] In contrast to the case of free exploration { \color{black}(see Sec.~\ref{sec:freeexpl})}, the second moment now depends on the full distribution $\psi(\mathcal{T})$ rather than 
    { \color{black}its} mean { \color{black}[see Eq.~\eqref{eq:secmomfree}]}.
    \item [ii)] In the limit of weak resetting $a\to 1$, we recover the steady-state mean squared displacement $D\tau_{\rm rel}$ of a Brownian particle in a harmonic trap as expected, since no resetting happens in this limit. In contrast to the case of free exploration { \color{black}[see Eq.~\eqref{eq:secmomfree}]}, the variance no longer diverges. { \color{black} In fact, since the background potential confines the particle trajectories, this result makes sense also for $a>1$ to a certain extent. From Eq. (\ref{eq:Vx2}) we see a that a finite steady state value only is found as long as $a < \sqrt{1/ \tilde \psi(2/\tau_\text{rel})}$,  which is the range of $a$-values for which the harmonic background potential is still able to confine the particle trajectories. In the absence of a potential, the dynamics of such diffusive processes with sudden state-dependent jumps has been studied in Refs. \cite{harbola2023stochastic,harbola2023stochastic2}. }
    \item [iii)] In the limit of vanishing potential{ \color{black}~$V(x)$ (i.e., $\tau_{\rm rel}\to \infty$),} the limit must be taken by first expanding $\tilde \psi(2/\tau_{\rm rel})$ for small arguments; $\tilde \psi(2/\tau_{\rm rel}) = 1 - \frac{2}{\tau_{\rm rel}}\langle \mathcal{T}\rangle+...$. 
    { \color{black}This recovers} the second moment $2D \langle \mathcal{T}\rangle/(1-a^2)$ { \color{black}as observed in Eq.~\eqref{eq:secmomfree}}. 
\end{itemize}

\section{Thermodynamic work}\label{sec:work}

Since resetting gives rise to non-equilibrium steady states, it is interesting to characterize exactly how far from equilibrium they are. This can be characterized by a thermodynamic cost, for example the entropy or work associated with the protocol $\lambda(t)$ that determines how the total potential in Eq. (\ref{eq:totpot}) is varied. Work performed on the system gives it energy, which can subsequently be dissipated into the heat bath (entropy is produced). Work and entropy are simply related through the first law of thermodynamics, and differ only by the internal energy difference at the beginning and end of the protocol. 

In overdamped systems where a potential is varied externally using a protocol $\lambda(t)$, stochastic thermodynamics associates a work given by Eq. (\ref{eq:thework}). Since the potential only changes at the time instances $t_k$ associated with the transition from exploration to resetting phase (or \emph{vice versa}), it is natural to consider the work in discrete time units $n$ that label the number of resets rather than fixing the total observation time.  The work associated with $n$ resets reads
\begin{eqnarray}
    w_n = \sum_{j=0}^n \left[\Phi(x^E_j) - V(x^E_j)\right] + \sum_{j=0}^n \left[ V(x^R_j) - \Phi(x^R_j)\right]\ ,
\end{eqnarray}
with $x^{E,R}_j$ being the position at the end of the $j$'th exploration ($E$) and resetting phases ($R$) as before. Since we are considering partial resets, we can use the fact that $x^R_n = a x^E_n$. In this case the work can be expressed only in terms of the position at the end of the $n$'th exploration phase:
\begin{eqnarray}\label{eq:work1}
    w_n = \sum_{j=0}^n \left[\Phi(x^E_j) - \Phi(a x^E_j) \right] + \sum_{j=0}^n \left[ V(a x^E_j) - V(x^E_j) \right]\ .
\end{eqnarray}
In the limit of weak resetting $a\to 1$ we see that the work vanishes identically, as expected. 
{ \color{black}In the following section, we} investigate how the mean work depends on the resetting strength in different resetting traps.

\subsection{Rate of work with free exploration.}
Consider a case where the resetting is mediated by a potential of the form 
\begin{eqnarray}
    \Phi(x) = \frac{1}{\zeta}{ \color{black}\kappa_R} |x|^\zeta\ ,
\end{eqnarray}
where $\zeta\in \mathbf{N}$ determines the shape of the resetting trap and ${ \color{black}\kappa_R}$ its strength. When there is no exploration potential $V(x) =0$ the work in Eq. (\ref{eq:work1}) reads
\begin{eqnarray}\label{eq:anharm}
    \langle w_n\rangle  = \frac{{ \color{black}\kappa_R}}{\zeta}  \sum_{j=0}^n \left[ \langle |x^E_j|^\zeta\rangle -\langle |x^R_j|^\zeta\rangle \right] = \frac{{ \color{black}\kappa_R}}{\zeta}  (1-a^\zeta)\sum_{j=0}^n  \langle |x^E_j|^\zeta\rangle \ .
\end{eqnarray}
Note that there is also an $a$ dependence coming from the moment itself. Below we consider the mean work in the case of two different resetting potentials.

\begin{figure}
    \centering
    \includegraphics[width = 14cm]{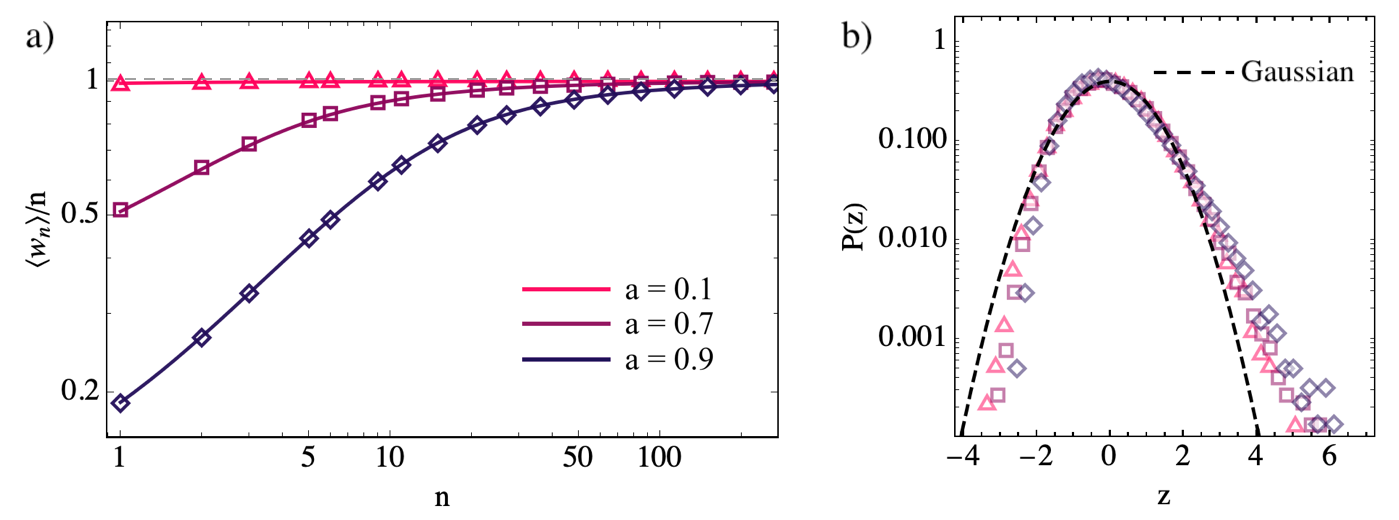}
    \caption{a) Work as a function of $n$ for various resetting strengths $a$, in the case of a harmonic resetting potential and $\psi(\mathcal{T}) = r \exp(-r \mathcal{T})$. Solid line shows theory, Eq. (\ref{eq:harmonicsolution}), while dots are { \color{black}the} numerical { \color{black}simulation} data. Dashed { \color{black} gray} line shows the asymptotic work rate { \color{black} $\kappa_R D\langle \mathcal{T}\rangle=1$.} { \color{black} b) Probability density for the rescaled work, $z \equiv (w_n-\langle w_n\rangle)/\sqrt{\text{Var}(w_n)}$, obtained using numerical simulations. Dashed curve shows a Gaussian distribution with unit variance and zero mean.} Parameters used are ${ \color{black}\kappa_R} = D = \gamma = r = 1$.}
    \label{fig:w1}
\end{figure}

\subsubsection{Harmonic resetting potential ($\zeta = 2$).}
Using Eq. (\ref{eq:anharm}) we have in the harmonic case
\begin{eqnarray}
    \langle w_n\rangle  = \frac{{ \color{black}\kappa_R}}{2}  (1-a^2)\sum_{j=0}^n  \langle (x^E_{ \color{black} j})^2 \rangle\ . 
\end{eqnarray}
{ \color{black}The second moment of position} we know from earlier sections { \color{black}[see Eq.~\eqref{eq:secmomfree}]}, which gives
\begin{eqnarray}
    \langle w_n\rangle  = { \color{black}\kappa_R}  D \langle \mathcal{T}\rangle  \sum_{j=1}^n  (1-a^{2j})\ .
\end{eqnarray}
The sum can be performed explicitly, resulting in the work
\begin{eqnarray}\label{eq:harmonicsolution}
    \langle w_n\rangle  =  { \color{black}\kappa_R}   D \langle \mathcal{T}\rangle  \frac{a^{2 n+2}-a^2 (n+1)+n}{1-a^2}\ .
\end{eqnarray}

{ \color{black}Figure~\ref{fig:w1}a} compares this result~\eqref{eq:harmonicsolution} 
{ \color{black}with numerical Langevin} simulations. At large $n$, corresponding to the steady state regime, the rate of work 
{ \color{black}becomes}
\begin{eqnarray}\label{asym-wrk}
    \lim_{n\to \infty} \frac{\langle w_n\rangle}{n}= \lim_{n\to \infty} [\langle w_{n+1}\rangle-\langle w_n\rangle] = { \color{black}\kappa_R} D \langle \mathcal{T}\rangle\ .
\end{eqnarray}

Interestingly, the asymptotic rate of work{ \color{black}~\eqref{eq:harmonicsolution}} does not depend on the resetting strength $a$. While the position distribution at the end of the exploration phases strongly depended on the { \color{black} resetting strength $a$} and underwent a transition from non-Gaussian to Gaussian as $a$ was varied across  { \color{black}$[0,1)$}, the long-time mean rate of work is invariant to such changes as long as $a$ is strictly less than $1$; at $a=1$ the work vanishes. {  \color{black} Interestingly, this work coincides with the work associated with a protocol where a harmonic potential is intermittently turned on and off for a random time \cite{olsen2023thermodynamic, alston2022non}. }

{ \color{black}We also perform numerical simulations to obtain the full steady state distribution of the work, as shown in Fig.~\ref{fig:w1}b. While the mean is invariant 
with respect to the resetting strength $a$, the probability density 
shows a weak dependence on the value of $a$ in its tails ($|z|\gg 1$). Moreover, the distribution is more skewed towards positive values of the work than a symmetric Gaussian. Typical fluctuations appear to be Gaussian however.}

\subsubsection{Anharmonic resetting potential ($\zeta = 4$).}
\begin{figure}
    \centering
    \includegraphics[width = 9cm]{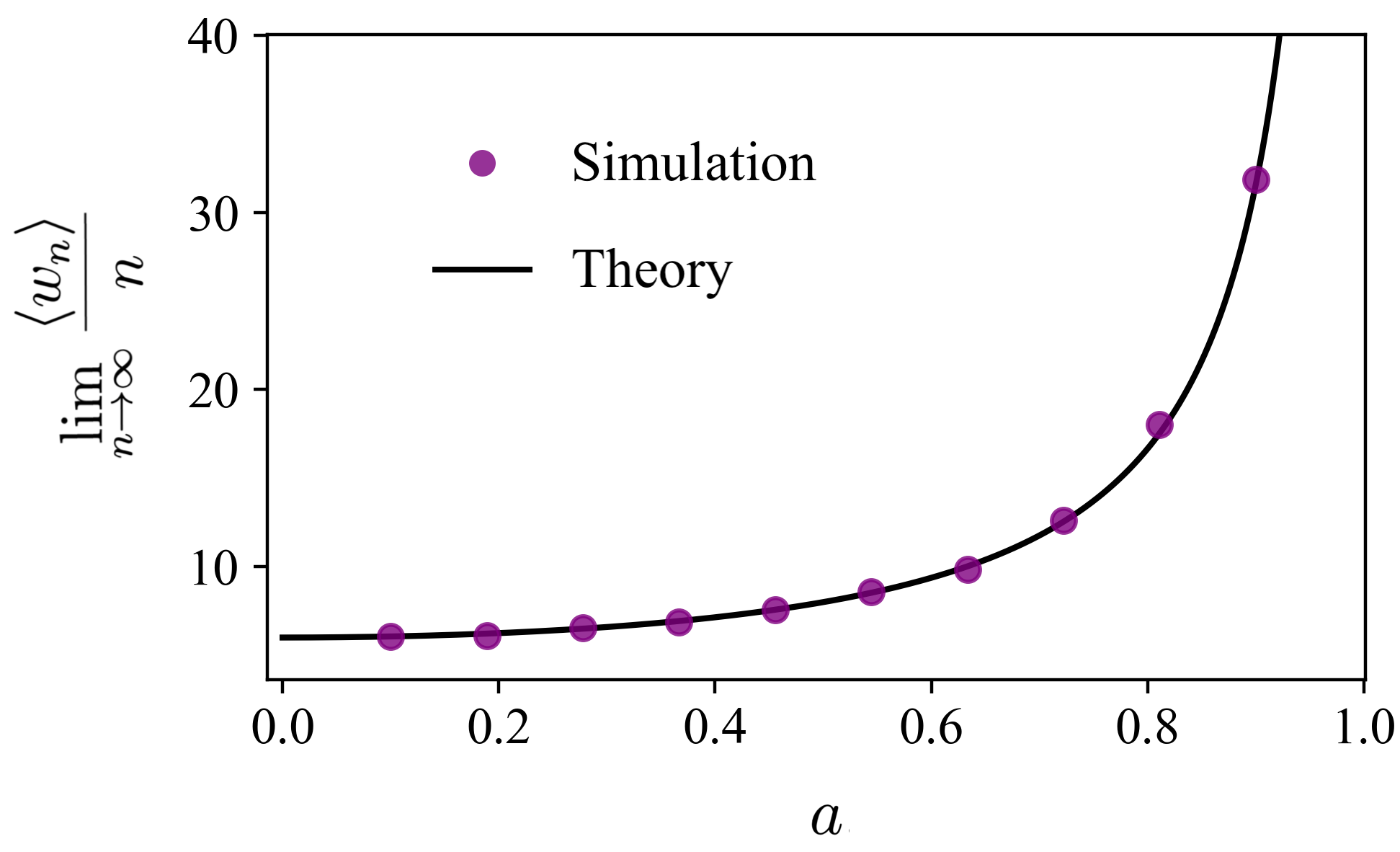}
    \caption{Long-time rate of work as a function of resetting strength $a$, in the case of anharmonic resetting potential and $\psi(\mathcal{T}) = r \exp(-r \mathcal{T})$. Solid line shows theoretical prediction from Eq. (\ref{eq:z4}), while dots are { \color{black}obtained from the~}numerical simulations. Parameters used are ${ \color{black}\kappa_R} = D = \gamma = r = 1$.}
    \label{fig:w2}
\end{figure}

Next we consider an anharmonic potential. Using Eq. (\ref{eq:anharm}) with $\zeta = 4$ we have the mean work
\begin{eqnarray}
    \langle w_n\rangle  = \frac{{ \color{black}\kappa_R}}{4}  (1-a^4)\sum_{j=0}^n  \langle (x^E_n)^4 \rangle\ . 
\end{eqnarray}
Rather than explicitly calculating the work for all $n$, we directly evaluate the steady-state rate of work, which is obtained by noting that
\begin{eqnarray}
    \langle w_n\rangle  \simeq \frac{{ \color{black}\kappa_R}}{4}  (1-a^4) n  \langle (x^E)^4 \rangle \ ,
\end{eqnarray}
where we used $\simeq$ to denote asymptotic equality at large $n$, and 
$\langle (x^E)^4 \rangle $ is the steady state fourth moment~{ \color{black}\eqref{eq:fourth}}. Combining with results from past sections, Eq. (\ref{eq:fourth}), we find 

\begin{eqnarray}\label{eq:z4}
    \lim_{n\to \infty} \frac{\langle w_n\rangle}{n} = \frac{{ \color{black}\kappa_R}}{4}\left[ 12 D^2 \langle \mathcal{T}^2\rangle  +  a^2  \frac{24 (D \langle \mathcal{T}\rangle)^2 }{1-a^2}\right]\ .
\end{eqnarray}
In contrast to the case of the harmonic potential~{ \color{black}\eqref{asym-wrk}}, here there is a clear dependence on the resetting strength $a$. Interestingly, the rate of work is monotonically growing as a function of $a$ [see Fig.~\ref{fig:w2}], while we simultaneously know from general arguments { \color{black}[see from Eq.~\eqref{eq:work1}]} that the work vanishes at $a=1$. Hence, it seems to be a sharp transition close to $a=1$, where the work jumps discontinuously from zero at $a=1$ to a value determined by the shape of $\Phi(x)$ for $a < 1$.

\subsubsection{Leading-order dependence on resetting strength for general anharmonic resetting traps.}

While it is hard to derive exact expressions for the asymptotic rate of work for potentials with exponent $\zeta$ for all $a$, it is possible to make general statements regarding growth or decay. We have previously seen that for the case of a harmonic resetting trap{ \color{black}[see Eq.~\eqref{asym-wrk}]}, the late-time work rate is invariant with respect to $a$, which was not the case for $\zeta = 4$ { \color{black}[see Eq.~\eqref{eq:z4}]}. Here, we extend these results by studying the leading-order behavior of the asymptotic work rate at small values of $a$.

First, we note that from Eq. (\ref{eq:anharm}) we have
\begin{eqnarray}
    \lim_{n\to \infty} \frac{\langle w_n\rangle}{n}  =  \frac{{ \color{black}\kappa_R}}{\zeta}  (1-a^\zeta) \langle |x^E|^\zeta\rangle\ .
\end{eqnarray}
 Multiplying Eq. (\ref{eq:rhoEfinal}) with $|x^E_n|^\zeta$ and integrating, we find 
\begin{equation}
    \langle (x^E_n)^\zeta \rangle=  \int d[x^E_{n-1}] \int d\mathcal{T}\psi(\mathcal{T})  \langle |x^E_n|^\zeta| a x^E_{n-1} \rangle_E(\mathcal{T})  \rho_{n-1}\left(x^E_{n-1}\right)\ .
\end{equation}
For free exploration the propagator is Gaussian, and we have 
\begin{eqnarray}
    \langle |x^E_n|^\zeta| a x^E_{n-1} \rangle_E(\mathcal{T}) &= (4 D \mathcal{T})^{\zeta/2} \frac{\Gamma\left(\frac{1+\zeta}{2}\right)}{\sqrt{\pi}} {\:}_1F_1\left(-\frac{\zeta}{2}, \frac{1}{2},- \frac{1}{2}  \left[ \frac{(ax^E_{n-1} )^2}{2D \mathcal{T}}\right]  \right)\nonumber \\
    & \approx (4 D \mathcal{T})^{\zeta/2} \frac{\Gamma\left(\frac{1+\zeta}{2}\right)}{\sqrt{\pi}} \left( 1 + \zeta  \frac{1}{2}  \left[ \frac{(ax^E_{n-1} )^2}{2D \mathcal{T}}\right] \right)\ ,
\end{eqnarray}
where ${\:}_1F_2(x,y,z)$ is the confluent hypergeometric function of the first kind, for which we used the series expansion to second order in resetting strength $a$. Since we are interested in the large $n$ behavior, we ignore the $n$ indices:
\begin{eqnarray}
    \langle |x^E|^\zeta| a x^E \rangle_E(\mathcal{T}) \approx (4 D \mathcal{T})^{\zeta/2} \frac{\Gamma\left(\frac{1+\zeta}{2}\right)}{\sqrt{\pi}} \left( 1 + \zeta  \frac{1}{2}  \left[ \frac{(ax^E )^2}{2D \mathcal{T}}\right] \right)\ .
\end{eqnarray}
Hence, we find
\begin{eqnarray}
    \langle |x^E|^\zeta\rangle  \approx (4 D)^{\zeta/2} \langle \mathcal{T}^{\zeta/2}  \rangle \frac{\Gamma\left(\frac{1+\zeta}{2}\right)}{\sqrt{\pi}} \left( 1 +   \frac{\zeta}{2}   \frac{a^2}{1-a^2} \right)\ ,
\end{eqnarray}
where we used the steady-state value for the second moment of $x^E$ from section \ref{sec:moments}. 
{ \color{black}Ignoring pre-factors, we find a work rate that scales as} 
\begin{eqnarray}
    \lim_{n\to \infty} \frac{\langle w_n \rangle}{n} &\sim (1-a^\zeta)\left( 1 +   \frac{\zeta}{2}   \frac{a^2}{1-a^2} \right)\ ,\\
    & \sim (1-a^\zeta)\left( 1 +   \frac{\zeta}{2}  a^2 \right) \ ,
\end{eqnarray}
for small $a$, where we have only kept terms up to second order. First, we note that the leading-order dependence on $a$ depends on the value of the potential exponent $\zeta$. In particular, for large $\zeta > 2$ the leading order is given by the right-most parenthesis, while for $\zeta < 2$ the first parenthesis is leading. Keeping only leading-order scaling (within our $\mathcal{O}(a^2)$ approximation), we find 
\begin{eqnarray}\label{asymptotic}
    \lim_{n\to \infty} \frac{\langle w_n \rangle}{n} &\sim
     \begin{cases}
      1-a^\zeta, & \textnormal{for } 0 < \zeta < 2,\\
      1, & \textnormal{for }\zeta = 2\ ,\\
       1 +   \frac{\zeta}{2}  a^2\ ,  &\textnormal{for } \zeta > 2\ .\\
     \end{cases}
\end{eqnarray}
We see that the case of a harmonic resetting trap ($\zeta = 2$) separates between the case of decaying ($\zeta <2$) and growing ($\zeta > 2$) work rates as a function of reset strength $a$. As always, the work rate vanishes exactly at $a=1$ independently of the chosen resetting trap{ \color{black}~[see from Eq.~\eqref{eq:work1}]}.

\subsection{Work in the presence of a background exploration potential}\label{sec:withpot}

In the past section, we have seen that the large-$n$ rate of work for a diffusive system with free exploration phases has some surprising properties as a function of the resetting strength { \color{black}$a$}. For harmonic resetting traps, the rate of work was independent of the resetting strength{ \color{black}~\eqref{asymptotic}}. Here, we show that this is no longer the case when a static background potential is present in the exploration phase. 

\begin{figure}
    \centering
    \includegraphics[width = 14cm]{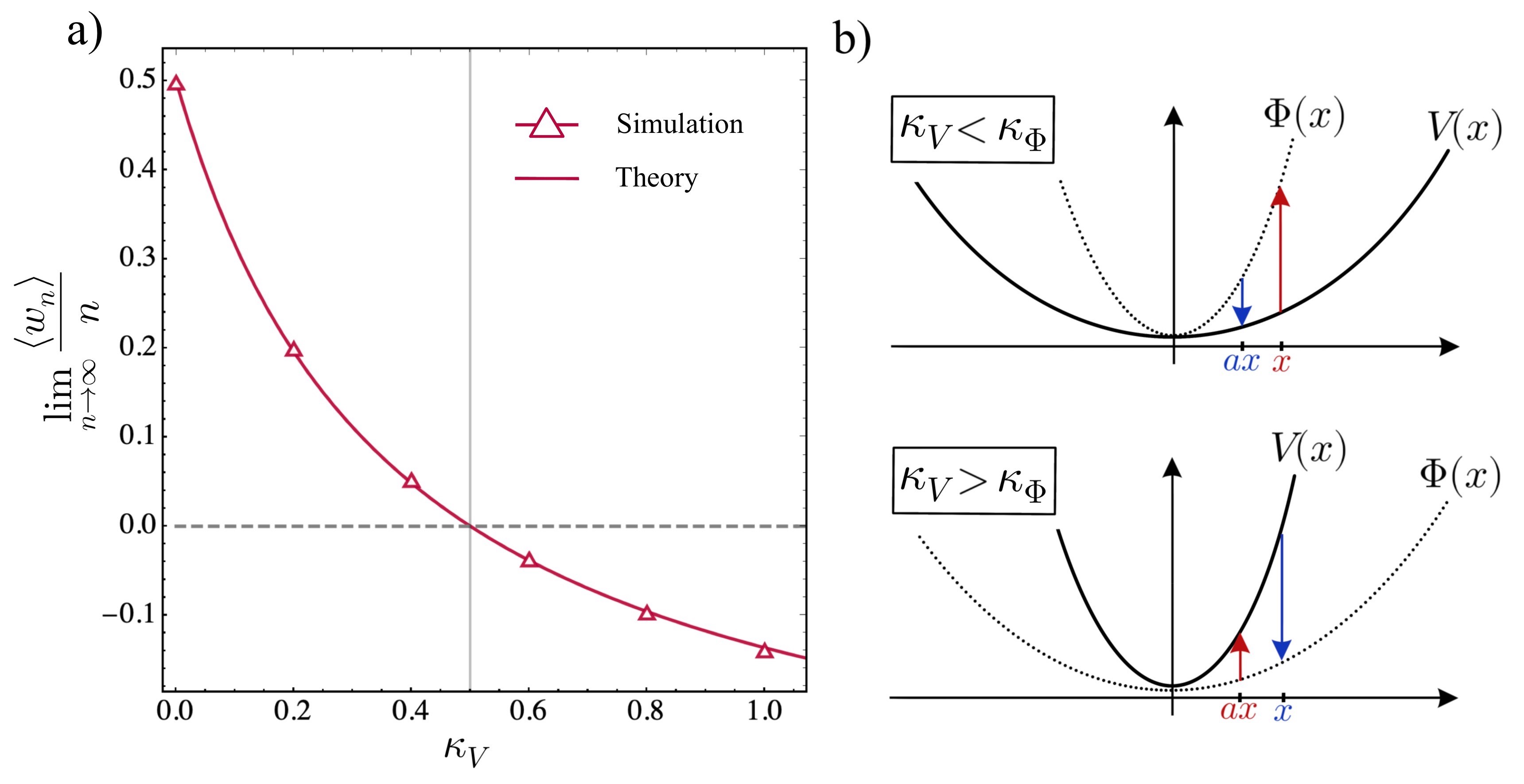}
    \caption{a) { \color{black}Large-$n$} rate of work for a diffusion process with partial resetting, with a harmonic exploration potential and a harmonic resetting trap. Points show 
    { \color{black}the numerical simulation} data, while solid line shows 
    { \color{black}the analytical result} [Eq. (\ref{eq:withV})]. { \color{black}Horizontal da}shed line is included at zero rate of work as a guide { \color{black}to the eye}. Vertical line is placed at ${ \color{black} \kappa_V} = { \color{black} \kappa_\Phi} = 1/2$. Parameters used are $D = r = \gamma = 1$, and $a = 1/2$. b) Sketch of the energetic cost of a single resetting event  for ${ \color{black} \kappa_V} < { \color{black} \kappa_\Phi}$ [top panel] and ${ \color{black} \kappa_V} > { \color{black} \kappa_\Phi}$ [bottom panel], elucidating why the work rate can become negative; for stiffer resetting traps, a higher price is payed for the switch $V(x) \to \Phi(x)$ [red arrow] than what is returned upon $\Phi(x)\rightarrow V(x)$ [blue arrow]. For stiffer exploration potentials, the opposite is true.  }
    \label{fig:withpot}
\end{figure}
We consider the simplest case, where the resetting is mediated by the trap
\begin{eqnarray}
    \Phi(x) = \frac{1}{2} { \color{black} \kappa_\Phi} x^2\ ,
\end{eqnarray}
while in the exploration phase another harmonic trap is present 
\begin{eqnarray}
    V(x) = \frac{1}{2} { \color{black} \kappa_V} x^2\ .
\end{eqnarray}
Using Eq. (\ref{eq:work1}), the steady state rate of work can be calculated as 
\begin{eqnarray}\label{sswrk}
    \lim_{n\to\infty} \frac{\langle w_n\rangle}{n} = \frac{{ \color{black} \kappa_\Phi} -{ \color{black} \kappa_V}}{2} (1-a^2)  \langle (x^E_j)^2\rangle\ ,
\end{eqnarray}
with $\langle (x^E_j)^2\rangle$ given by  Eq. (\ref{eq:Vx2}). 
{ \color{black}Then, the above equation~\eqref{sswrk} becomes:}
\begin{eqnarray}
    \lim_{n\to\infty} \frac{\langle w_n\rangle}{n} = \frac{{ \color{black} \kappa_\Phi}-{ \color{black} \kappa_V}}{2} (1-a^2) D \tau_{\rm rel} \frac{1-\tilde \psi(2/\tau_{\rm rel})}{1-a^2 \tilde \psi(2/\tau_{\rm rel})}\ ,
\end{eqnarray}
where we recall that $\tau_{\rm rel} = \gamma/{ \color{black} \kappa_V}$. Since the Laplace transform of a probability density is a positive number smaller than unity, we see that the sign of this expression is determined only by the sign of $\left({ \color{black} \kappa_\Phi}-{ \color{black} \kappa_V}\right)$. The rate of work decays monotonically as a function of $a$ for ${ \color{black} \kappa_V} < { \color{black} \kappa_\Phi}$, i.e., resetting traps that are sharper than the exploration trap. For ${ \color{black} \kappa_V} > { \color{black} \kappa_\Phi}$, the rate of work is negative and grows (becomes less negative) as $a$ is increased. In either case, the work rate vanishes at $a=1$ { \color{black}~[see from Eq.~\eqref{eq:work1}]}. In contrast to the case of free exploration phase, the rate of work is now a continuous function of the resetting strength $a$, and the $a\to 1$ limit is no longer singular.

As a simple example, we consider exponentially distributed exploration phase durations with rate $r$, in which case $\tilde \psi(s) = r/(s+r)$. The steady state work rate then takes the form 
\begin{eqnarray}\label{eq:withV}
   \lim_{n\to\infty} \frac{\langle w_n\rangle}{n} = \left({ \color{black} \kappa_\Phi} - { \color{black} \kappa_V}\right)  \frac{\left(1-a^2\right) \gamma  D }{\left(1-a^2\right) \gamma  r+2 { \color{black} \kappa_V}}\ .
\end{eqnarray}

Figure~\ref{fig:withpot} summarizes the main properties of this case, and compares the analytical prediction 
{ \color{black}with the} numerical simulations.

\section{Discussion}\label{sec:concl}
Partial resetting has been considered within the framework of intermittent fluctuating potentials. In the fixed-$n$ ensemble, we have derived exact relations obeyed by the propagator constrained to the end-points of exploration and resetting phases. In strong contrast to other intermittent potential schemes used to model complete resetting, here we see that the results are invariant under changes to the resetting phase dynamics. For free exploration, we have shown that under general conditions there is a transition in the end-of-exploration density in the stationary regime, where at strong resetting it is determined by the details of the dynamics, while at weak resetting it always approaches a Gaussian. We used the steady-state results to calculate the thermodynamic work associated with performing the resets. For a harmonic potential, we have shown that the mean rate of work surprisingly is independent { \color{black}of} the strength of resetting. For other resetting traps, the resetting strength matters. Indeed, for small values of $a$ we have shown that for a potential of the form $\Phi(x) \sim |x|^\zeta$, the asymptotic rate of work grows as a function of $a$ for $\zeta > 2$, while it decays whenever $\zeta < 2$. For an anharmonic resetting trap $\Phi(x) \sim x^4$, the work rate diverges as the resetting strength becomes weaker. The rate of work was also calculated when a harmonic potential is present in the exploration phase, displaying a more involved dependence on the resetting strength. Numerical simulations confirmed our results. 

Here various results have been derived in the fixed-$n$ ensemble, where we only have information regarding the particle's position at the end of the $n$'th exploration phase. In the future, it could be interesting to address the fully time-resolved version of this problem, to see which properties are valid also in that case. Indeed, since the duration of the resetting depends both on the shape of the resetting trap and the resetting strength, we expect an ensemble of fixed observation time to have quite different properties than in the fixed-$n$ case. Both ensembles are natural in experiments however, and whether one chooses to fix number of resets or total observation time is a matter of preference. 

It would also be interesting to develop a thermodynamic theory that incorporates work fluctuations as was done recently for conventional resetting \cite{olsen2023thermodynamic}. Finally, it would be interesting to corroborate the predictions of this paper with experimental implementations using optical methods.

\begin{acknowledgements}
KSO acknowledges support by the Deutsche Forschungsgemeinschaft (DFG) within the project LO 418/29-1. DG acknowledges the Nordita fellowship program. Nordita is partially funded by Nordforsk. 
\end{acknowledgements}


\appendix

\section{Weak resetting limit}
Here we show that there is a transition in the steady state from non-Gaussian to Gaussian as $a\to 1$ for the propagator constrained to the endpoint of exploration phases for any distribution $\psi(\mathcal{T})$. Following Ref. \cite{tal2022diffusion}, we consider the scaled density introduced in section \ref{sec:freeexpl}, i.e. Eq. (\ref{eq:Pz}). In the limit $a\to 0$ only the $j=0$ factor contributes in Eq. (\ref{eq:gen}), and we simply have
\begin{eqnarray}\label{eq:stronk}
    \hat P(k;0) = \int d\mathcal{T}\psi(\mathcal{T})\hat G_E(k/\Sigma(0),\mathcal{T})\ .
\end{eqnarray}
Here, we have used the transformation properties of a probability density and its Fourier transform. Conversely, in the limit of very weak resetting $a\to 1$ we have
\begin{eqnarray}
    \hat P(k;1) = \lim_{a\to 1}\prod_{j=0}^\infty  \left[ \int d\mathcal{T}\psi(\mathcal{T})\hat G_E\left(\frac{a^jk}{\Sigma(a)},\mathcal{T}\right) \right] \ .
\end{eqnarray}
Since for $a\to 1$ we have $\frac{a^jk}{\Sigma(a)} \to 0$, the Fourier transform can be Taylor expanded \cite{tal2022diffusion}. To second order we have
\begin{eqnarray}
    \hat P(k;1) = \lim_{a\to 1}\prod_{j=0}^\infty  \left[1 -\frac{1}{2} \left[ \frac{a^jk}{\Sigma(a)} \right]^2\int d\mathcal{T}\psi(\mathcal{T})\sigma^2(\mathcal{T}) \right] \ .
\end{eqnarray}
This type of product is well known to approach an exponential, and we are left with
\begin{eqnarray}
    \hat P(k;1) = \lim_{a\to 1} \exp\left( - \frac{\Sigma(0) k^2}{2 (1-a^2)\Sigma^2(a)} \right) =\exp\left( - \frac{k^2}{2} \right)\ .  
\end{eqnarray}
This is simply the Fourier transform of a Gaussian with unit variance and zero mean. Hence, the transition as resetting strength is varied is very general, and is expected to hold for a very broad class of systems; at small values of $a$, we have a density given by Eq. (\ref{eq:stronk}), while at larger values of $a$ the density transitions to a Gaussian. It is worth emphasizing that here we have not assumed much about the underlying propagator of the system, except for spatial homogeneity, normalizability, and that its Fourier transform can be expanded at least to second order. We therefore expect this transition to be rather general and also hold for non-Brownian systems. We note however, that here the transition is shown to hold for the stationary regime of the end-of-exploration density, while in Ref. \cite{tal2022diffusion} the transition was observed in the fully time-resolved steady state.


%

\end{document}